\documentclass[manuscript]{aastex}

\slugcomment{Not to appear in Nonlearned J., 45.}

\shorttitle{The Patterns of High Level Magnetic Activity of V1285 Aql}
\shortauthors{Dal and Evren}

\begin{document}


\title{The Patterns of High Level Magnetic Activity \\
Occurring on the Surface of V1285 Aql:\\
The OPEA Model of Flares and DFT Models of Stellar Spots}


\author{H. A. Dal and S. Evren}
\affil{Department Of Astronomy and Space Sciences, University of Ege, \\
Bornova, 35100 ~\.{I}zmir, Turkey}

\email{ali.dal@ege.edu.tr}



\begin{abstract}
Statistically analysing Johnson UBVR observations of \object{V1285 Aql} during the three observing seasons, both activity level and behaviour of the star are discussed in respect to obtained results. We also discuss the variation out-of-flare due to rotational modulation. 83 flares were detected in U band observations of the season 2006. First of all, depending on statistical analyses using the Independent Samples t-Test, the flares were divided into two classes as the fast and the slow flares. According to the results of the test, there is a difference of about 73 s between the flare-equivalent durations of slow and fast flares. The difference should be the difference mentioned in the theoretical models. Secondly using the One Phase Exponential Association function, the distribution of the flare-equivalent durations versus the flare total durations was modelled. Analysing the model, some parameters such as $Plateau$, $Half-Life$ values, the mean average of the flare-equivalent durations, maximum flare rise and total duration times are derived. The $Plateau$ value, which is an indicator of the saturation level of white-light flares, was derived as 2.421$\pm$0.058 s in this model, while $Half-Life$ is computed as 201 s. Analyses showed that observed maximum value of flare total duration is 4641 s, while observed maximum flare rise time is 1817 s. According to these results, although computed energies of the flares occurring on the surface of \object{V1285 Aql} are generally lower than those of other stars, the length of its flaring loop can be higher that those of more active stars. Moreover, the variation out-of-flare activity was analysed with using three methods of time series analysis, a sinusoidal-like variation with period of $3^{d}.1265$ was found for rotational modulation out-of-flare for the first time in literature. Considering the variations of V-R colour, these variations must be because of some dark spot(s) on the surface of that star. In addition, using the ephemeris obtained from time series analyses, the distribution of the flares was examined. The phase of maximum mean flare occurrence rates and the phase of rotational modulation were compared to investigate whether there is any longitudinal relation between stellar flares and spots. The analyses show that there is a tendency of longitudinal relation between stellar flares and spot(s). Finally, it was tested whether slow flares are the fast flares occurring on the opposite side of the stars according to the direction of the observers as mentioned in the hypothesis developed by \citet{Gur86}. The flare occurrence rates reveal that both slow and fast flares can occur in any rotational phases.

\end{abstract}





\section{Introduction}

Flares and flare processes observed on the surfaces of UV Ceti type stars have not been perfectly understood yet, although they are heavily studied subjects of astrophysics \citep{Ben10}. In this study, we obtained large data set in UBVR bands from the observations of \object{V1285 Aql}. Depending on these large photometric data, which is very useful for a statistical analysis of the flare properties, we have obtained some remarkable results. Observed star, \object{V1285 Aql}, is classified as a UV Ceti type star from spectral type dM3e in SIMBAD data base. According to \citet{Vee74}, the star seems to belong to the young disk population of the galaxy and is classified as a young flare star. The flare activity of \object{V1285 Aql} was discovered for the first time by \citet{Sha70}. Apart from flare activity, \citet{And88} showed that \object{V1285 Aql} exhibits some sinusoidal-like variations out-of-flare with period of 30 s. However, later it was found that the star exhibits the same variations with period of 1.2 and 1.4 minutes \citep{And89}. In fact, it is a debate issue whether \object{V1285 Aql} exhibits any rotational modulation, or not. Moreover, the period of the equatorial rotation is found to be $2^{d}.9$, this is an other debate issue for \object{V1285 Aql} \citep{Doy87, Ale97, Mes01}.

\object{V1285 Aql} was observed in U band for flare patrol in 2006, and 83 white-light flares were detected in U band. Considering the studies of \citet{Har69, Osa68, Mof74, Gur88} and following the method described by \citet{Dal10}, first of all, we analysed large U band flare data in order to classify flares. This is because the classification of the flare light variations is important due to modelling the event \citep{Gur88, Ger05}. In the literature, white-light flare events observed on the surfaces of UV Ceti type stars were usually classified into two types as slow and fast flares \citep{Har69, Osa68}. On the other hand, both \citet{Osk69} and \citet{Mof74} classified flares in more than two types. According to Kunkel, the observed flare light variations should be a combination of slow and fast flares \citep{Ger05}. Finally \citet{Dal10} developed a rule, which is depending on the ratios of flare decay times to flare rise times. According to \citet{Dal10}, if the decay time of a flare is 3.5 times longer than its rise time at least, the flare is a fast flare. If not, the flare is a slow flare. They demonstrated that the value of 3.5 is a boundary limit between the two types of flare.

In the second step, we analysed flare data set to find general properties of flare events occurring on the surface of \object{V1285 Aql}. The method described by \citet{Dal11a} was followed for these analyses. The energy limit and some timescales of the flare events occurring on a star are as important as the types of these event. In the literature, \citet{Gur88} stated about two processes as thermal and non-thermal processes, and mentioned that there must be a large energy difference between these two types of flares. Moreover, \citet{Ger72, Lac76, Wal81, Ger83, Pet84} and, \citet{Mav86} studied on the distributions of flare energy spectra of UV Ceti type stars. There are significant differences between energy levels of stars from different ages. Depending on the processes of Solar Flare Event, however, flare activity seen on the surfaces of dMe stars is generally modelled. This is why the magnetic reconnection process is accepted as the source of the energy in these events \citep{Ger05, Hud97}. According to both some models and observations, it is seen that some parameters of magnetic activity can reach the saturation \citep{Ger05, Sku86, Vil83, Vil86, Doy96a, Doy96b}. Recently, \citet{Dal11a} have been examined the distributions of flare equivalent durations versus flare total durations. In the analyses, the distributions of flare-equivalent durations were modelled by the One Phase Exponential Association function (hereafter the OPEA). In the models, it is seen that flare-equivalent durations can not be higher than a specific value and it is no matter how long the flare total duration is. According to \citet{Dal11a}, this level, the $Plateau$ parameter, is an indicator for the saturation level of the flare process occurring on the surface of the program stars in some respects. In fact, white-light flares are detected in some large active regions such as compact and two-ribbon flares occurring on the surface of the Sun \citep{Rod90, Ben10}. It is possibly expected that the energies or the flare-equivalent durations of white-light flares can reach the saturation.

Moreover, it is well known that some samples of UV Ceti stars exhibit sinusoidal-like variations out-of-flare activity. The stars such as, EV Lac, V1005 Ori are well known samples \citep{Dal11b}. In this respect, apart from the flare patrol, \object{V1285 Aql} was observed in BVR bands from 2006 to 2008 in order to examine whether there is any sinusoidal-like variation due to rotational modulation. The variation out-of-flare activity was analysed with using three different methods of the time series analyses. In fact, a sinusoidal-like variation due to rotational modulation was found out-of-flare activity.

\section{Observations and Analyses}

\subsection{Observations}

The observations were acquired with a High-Speed Three Channel Photometer attached to the 48 cm Cassegrain type telescope at Ege University Observatory. Observations were grouped in two schedules. Using a tracking star in second channel of the photometer, flare observations were only continued in standard Johnson U band with exposure times between 7 and 10 seconds. The second observation schedule was used for determining whether there is any variation out-of-flare. Pausing flare patrol of program stars, we observed the star once or twice a night, when they were close to the celestial meridian. The observations in this schedule were made with the exposure time of 10 seconds in each band of standard Johnson BVR system, respectively. These observations were continued from the season 2006 to 2008. The same comparison stars were used for both types of observations.

Some properties of \object{V1285 Aql} and its comparisons are listed in Table 1. Standard V magnitudes and B-V colour indexes obtained in this study are given in Table 1. Although \object{V1285 Aql} and its comparison stars are very close to one another on the celestial plane, differential extinction corrections were applied. The extinction coefficients were obtained from observations of the comparison stars on each night. Moreover, the comparison stars were observed with the standard stars in their vicinity and the reduced differential magnitudes, in the sense of variable minus comparison stars, were transformed to the standard system using the procedures described by \citet{Har62}. The standard stars were chosen from the catalogues of \citet{Lan92}. Heliocentric corrections were applied to the times of the observations. The standard deviations of observation points acquired in the standard Johnson UBVR bands are about $0^{m}.015$, $0^{m}.009$, $0^{m}.007$ and $0^{m}.007$ on each night, respectively. To compute the standard deviations of observations, we used the standard deviations of the reduced differential magnitudes in the sense comparisons (C1) minus check (C2) stars for each night. There is no variation in the standard brightness of the comparison stars. The flare patrol of \object{V1285 Aql} was continued for 32 nights between May 5, 2006 and August 25, 2006. 83 flares were detected in this U band patrol.

\citet{Ger72} developed a method for calculating flare energies. Flare-equivalent durations and energies were calculated using Equations (1) and (2) of this method,

\begin{center}
\begin{equation}
P = \int[(I_{flare}-I_{0})/I_{0}] dt
\end{equation}
\end{center}
where $I_{0}$ is the intensity of the star in the quiescent level and $I_{flare}$ is the intensity during flare, and

\begin{center}
\begin{equation}
E = P \times L
\end{equation}
\end{center}
where $E$ is the flare energy, $P$ is the flare-equivalent duration, and $L$ is the luminosity of the stars in the quiescent level in the Johnson U band.

Some parameters such as HJD of flare maximum time, flare rise and decay times (s), amplitude of flare (mag), and flare-equivalent duration (s) were calculated for each flare. The procedure followed in the computations was described in detail by \citet{Dal10}. In brief, it is more important to note that the brightness of a star without a flare was taken as a quiescent level of the brightness of this star on each night. Considering this level, all flare parameters were calculated for each night. If a flare has a few peaks, both the maximum time and amplitude of the flare were calculated from the first highest peak. All calculated parameters are listed in Table 2 for 83 flares. The observing date, HJD of flare maximum time, flare rise and decay times (s), flare total duration (s), flare amplitude in U band (mag), U-B colour index (mag), flare energy (erg) and flare type are listed in the columns of the table, respectively. Although it was explained in detail by \citet{Dal10}, more important second note is that the flare-equivalent durations were used in the analyses due to the luminosity term ($L$) in Equation (2), in stead of flare energies.

When the observed flares are examined, it is seen that almost each flare has a distinctive light variation shape (Figures 1 - 4). In these figures, the horizontal dashed lines represent the level of quiescent brightness. According to the rule described by \citet{Dal10}, the flare shown in Figure 1 is a sample of fast flare. This type flare is classified as a flare event in the classification of \citet{Mof74}. According to \citet{Dal10}, the flares shown in Figure 2 and 4 are two samples of the fast flares, while these flares could be classified as classical flares by \citet{Mof74}, and the flare shown in Figure 3 is a complex flare sample.

\subsection{Fast and Slow Flares}

Following the method used by \citet{Dal10}, all detected flares were analysed to examine whether the limit ratio (3.50) is acceptable for the flares detected from \object{V1285 Aql}. Thus, using new-large data set, we have tested whether the value of 3.50 is a general limit, or not. First of all, in the analyses, we compared the equivalent durations of flares, whose rise times are equal. For instance, there are 16 flares, whose rise times are 15 s. The average of their equivalent durations is 15.026 s. Apart from these 16 flares, there are 5 other flares, whose rise times are similarly 15 s. However, the average of their equivalent durations is 5.454 s for these 5 flares. The main difference of these two example groups is seen in the shapes of the light curves. Finally, we found 48 flares with higher energy and 35 flares with lower energy among 83 flares detected from \object{V1285 Aql}.

In the analyses, the Independent Samples t-Test (hereafter t-Test) \citep{Wal03, Daw04} was used in the SPSS V17.0 \citep{Gre99} and GrahpPad Prism V5.02 \citep{Mot07} softwares in order to test whether these two groups are really independent from each other. The flare rise times were taken as a dependent variable in the t-Test, while the flare-equivalent durations were taken as an independent variable.  The value of ($\alpha$) is taken as 0.005, which gave us to test whether the results are statistically acceptable, or not \citep{Daw04}.

The mean averages of equivalent durations were compared for two groups. The mean average of the equivalent durations for 35 slow flares was calculated and found to be 1.479$\pm$0.054 s, and it was found to be 2.015$\pm$0.067 s for 48 fast flares in the logarithmic scale. This shows that there is a difference of about 0.536 between average equivalent durations in the logarithmic scale. The probability value (hereafter $p-value$) was computed to test the results of the t-Test, and it was found to be $p<0.0001$. Considering $\alpha$ value, this means that the result is statistically acceptable. All the results obtained from the t-Test analyses are given in Table 3.

In the second step, the distributions of the equivalent durations ($logP_{u}$) versus flare rise times ($logT_{r}$) were derived for both flare types. The best models for the distributions were searched. Using the Least-Squares Method, regression calculations showed that the best models of distributions are linear functions given by Equations (3) and (4). The derived linear fits are shown in Figure 5.

\begin{center}
\begin{equation}
log(P_{u})~=~1.150~\times~log(T_{r})~-~0.285
\end{equation}
\end{center}

\begin{center}
\begin{equation}
log(P_{u})~=~0.932~\times~log(T_{r})~+~0.385
\end{equation}
\end{center}

In the next step, it was tested whether these linear functions belong to two independent distributions, or not. At this point, the slopes of linear functions were principally examined. As can be seen in Table 3, the slope of the linear function is 0.932$\pm$0.056 for slow flares, while it is 1.150$\pm$0.095 for fast flares. This shows that the increase in equivalent durations versus flare rise times for both fast and slow flares is almost parallel. When the $p-value$ was calculated to test whether two fits can be statistically accepted as parallel, it was found to be $p=0.670$. This value indicates that there is no significant difference between the slopes of fits, and they are statistically parallel.

Finally, the y-intercept values of two linear fits were calculated and compared. While the $y-intercept$ value is -0.385 for the slow flares, it is -0.285 for the fast flares in the logarithmic scale. There is a difference of about 0.100 between them. When the $p-value$ was calculated for the $y-intercept$ values to say whether there is a statistically significant difference, it was found that $p<0.0001$. This result indicates that the difference between two $y-intercept$ values is clearly important.

To test whether there is a difference between maximum energy levels and timescales of the two flare types, the distributions of the equivalent durations in the logarithmic scale versus flare rise times were derived. The derived distributions are shown in Figure 6. Using the Least-Squares Method, regression calculations showed that the averaged value of upper limit is 2.928$\pm$0.251 for the fast flares, while it is 2.217$\pm$0.075 for the slow flares. Moreover, the lengths of flare rise times for both types of flares can be compared in Figure 6. While the lengths of rise times for slow flares can reach to 1817 s, they are not longer than 510 s for fast flares.

\subsection{The One-Phase Exponential Association Models of the Distribution of the Flares}

In order to test whether there are any upper limits for the distributions of the equivalent durations ($logP_{u}$), the distributions of the equivalent durations in the logarithmic scale versus flare total durations were examined. Using regression calculations, the best model fit was identified by SPSS V17.0 software for this distribution. Analyses showed that the OPEA function \citep{Mot07, Spa87} given by Equation (5) is the best model fit. According to \citet{Dal11a}, this is an expected case, and this demonstrated that the flares occurring on the surface of \object{V1285 Aql} have an upper limit for producing energy. Using the Least-Squares Method, the OPEA model of the distributions of the equivalent durations in the logarithmic scale versus flare total durations was derived by GrahpPad Prism V5.02.

\begin{center}
\begin{equation}
y~=~y_{0}~+~(Plateau~-~y_{0})~\times~(1~-~e^{-k~\times~x})
\end{equation}
\end{center}

Although the details of the OPEA function have been given by \citet{Dal11a}, in brief, there are some important parameters derived from this function, which are some indicators for the condition of the occurring flare processes. One of them is $y_{0}$, which is the lower limit of equivalent durations for observed flares in the logarithmic scale. In contrast to $y_{0}$, the parameter of $Plateau$ is the upper limit. The value of $y_{0}$ depends on the quality of observations as well as flare power, while the value of $Plateau$ depends only on power of flares. This parameter is identified as a saturation level for flare activity observed in U band by \citet{Dal11a}. The derived OPEA model is shown in Figure 7, while the parameters of the model are listed in Table 4. The $Span$ value listed in the table is the difference between the values of $Plateau$ and $y_{0}$. The $Half-Life$ value is half of the first $x$ values, where the model reaches the $Plateau$ values for a star. In other words, it is half of the flare total duration, where flares with the highest energy start to be seen. Moreover, statistical analyses showed that maximum flare rise time obtained from these 83 flares is 1817 s, while the maximum flare total duration is 4641 s.

On the other hand, using the t-Test, the $Plateau$ value derived from the model was tested whether the $Plateau$ value is statistically acceptable, or not. The flares in the $Plateau$ phases were only used to test. The mean average of the equivalent durations was computed and found to be 2.500$\pm$0.076. In fact, the $Plateau$ value had been found to be 2.421$\pm$0.058. Considering standard deviations of the parameters, these two parameters are almost equal to each other.

\subsection{The Rotational Modulation Out of Flares}

In order to determining whether there is any sinusoidal-like variation due to rotational modulation out-of-flare, we observed the star once or twice a night for three observing seasons. To purify the data from any flares or flare like variations, we used U band light and U-B colour as an indicators. This is because the U-B colour indexes are much more sensitive to the flare activity on the surface of the star. If a flare is too small to be detected in respect to the threshold, no sing is seen in V light, B-V and V-R colours. On the other hand, some distinctive sign could be seen in U band light and U-B colour. According to the results of these controls, some data and observations were disregarded for the analyses of sinusoidal-like variation.

After these eliminations, all the data sets in BVR bands were analysed with the method of Discrete Fourier Transform (DFT) \citep{Sca82}. The results obtained from DFT were tested by two other methods. One of them is CLEANest, which is another Fourier method \citep{Fos95}. The second method is the Phase Dispersion Minimization (PDM), which is a statistical method \citep{Ste78}. These methods confirmed the results obtained by DFT. Found photometric periods are 3.1269$\pm$0.0005 in B band, 3.1265$\pm$0.0005 in V band and 3.1268$\pm$0.0005 in R band. The photometric periods found from each sets are almost equal to each other.

Using the ephemeris given in Equation (6), which was found from V band by the DFT method \citep{Sca82}, the phases were computed for each season of three years. The obtained light curves in V band and colour curves of B-V and V-R are shown in Figure 8. 

\begin{center}
\begin{equation}
JD (Hel.) = ~24~53905.49106 + 3^{d}.1265~\times~E.
\end{equation}
\end{center}

From the DFT models of V band, the minimum phase of sinusoidal-like variation is $0^{P}.47$ in the season 2006, while it is $0^{P}.19$ for 2008 and $0^{P}.51$ for 2008. Although no variation in B-V colour curves is seen above the values of $3\sigma$ for each season, there are some variations in the V-R. If the colour curves of V-R are considered, it is seen that the star gets redden toward the minimum phase of the light curves in 2006 and 2008, while there is not any reddening or bluer in the V-R curve in 2007. Although it seems that there is a variation in the V-R curves of 2007, its amplitude is close to the values of $3\sigma$, and is almost lower than it. Considering the mean B-V value of $1^{m}.469$ and the variations in both light and colour curves, all these sinusoidal-like variation must be due to the rotational modulation of some region(s) covered by dark stellar spots.

\subsection{The Phase Distributions of the Flares}

The phases of flare maxima were computed for all flare types (together with fast and slow flares) with the same method used for the phases of light curves. The flare maximum times were used to compute the phases due to main energy emitting in this part of the flare light curves. In addition, the photometric periods of the stars are too long according to the average of flare total durations. In the second step, we computed the flare occurrence rates, the ratio of total flare number ($\Sigma n_{f}$) to total monitoring time ($\Sigma T_{t}$), in intervals of 0.10 phase length. Using Equation (7), we followed the method used by \citet{Let97} and \citet{Dal11b}.

\begin{center}
\begin{equation}
N~=~\Sigma n_{f}~/~\Sigma T_{t}
\end{equation}
\end{center}

The computed mean flare occurrence rates were plotted versus the rotational phase as a histogram. The plotted histogram is shown in Figure 9. Using the Least-Squares Method, the histogram of $N$ was analysed with the SPSS V17.0 \citep{Gre99} software to determine the phase in which Maximum Flare Occurrence Rates (hereafter MFOR) are seen. The analysis shows that the phase of MFOR is between the phases of $0^{P}.40$ - $0^{P}.50$ in the season 2006. This phase range is the same range of the minimum phase of sinusoidal-like variation due to rotational modulation found in the season 2006.

\citet{Gur86} developed a hypothesis called Fast Electron Hypothesis, in which red dwarfs generate only fast flares on their surface. On the other hand, according to the flare region on the surface of the star in respect to direction of observer, the shapes of the flare light variations can be seen like a slow flare. If the scenario in this hypothesis is working, it is expected that the fast and slow flares should be collected into two phases in the light curves of UV Ceti type stars, which exhibit BY Dra Syndrome. It is also expected that these two phases are separated from each other with intervals of $0^{P}.50$ in phase.

Then the phase distributions of fast flares were compared with the phases of slow flares in order to find out whether there is any separation as expected in this respect. When the phases of both fast and slow flares were examined one by one, it was clearly seen that both of them can occur in any phase. To reach a definite result, the phase distributions of both fast and slow flares were statistically investigated. The same method described above was used for these groups, too. The obtained histograms of both the fast and slow flares are shown in Figure 10. The histogram of the fast flares is shown in upper panel, while it is shown in bottom panel for the slow flares. These histograms were analysed with using the Least-Squares Method, the phase of MFOR was found to be $0^{P}.45$ for the fast flares, while it was found to be $0^{P}.30$ for the slow flares. The difference between the distributions of the two groups is about $0^{P}.15$.

\section{Results and Discussion}

\subsection{Flare Activity and Flare Types}

In this study, we detected 83 flares in U band observations of \object{V1285 Aql}. The samples of the fast and slow flares, whose rise times are equal, were determined as the suitable sets to analysis with using the t-Test. The results of the statistical t-Test analyses show that there are some distinctive differences between two data sets. The slope of the linear fit is 0.932 for slow flares, which are low energy flares, and it is 1.150 for fast flares, which are high energy flares. The values are almost close to each other. It demonstrates that the flare-equivalent durations versus the flare rise times increase in similar ways for both groups.

When the averages of equivalent durations for two types of flares were computed in the logarithmic scale, it was found that the average of equivalent durations is 1.479 for slow flares, and it is 2.205 for fast flares. The difference of 0.536 between these values in the logarithmic scale is equal to the 73.384 s difference between the equivalent durations. As can be seen from Equation (2), this difference between average equivalent durations affects the energies in the same way. Therefore, there is a difference of 73.384 between the energies of these two types of flares. This difference must be the difference mentioned by \citet{Gur88}. On the other hand, according to \citet{Dal10}, this difference between two flare types is about 157 s. Although the value found in this study is about half of the value found by \citet{Dal10}, but there are several reasons for this difference. The value given by \citet{Dal10} is an average derived from five flare stars. Moreover, there are EV Lac and EQ Peg among these five stars in the study of \citet{Dal10}. These two stars could cause some difference on the values, because they are seen to be dramatically different among other stars, as seen in Figures 5 and 6 given by \citet{Dal11a}. There must be some differences in the flare processes occurring on the surfaces of these stars.

Comparing the $y-intercept$ values of the linear fits, it is seen that there is a 0.100 times difference in the logarithmic scale, while there is a 0.536 times difference between general averages. Also considering Figure 5, it is seen that equivalent durations of fast flares can increase more than the equivalent durations of slow flare toward long rise times. Some other effects should be involved in the fast flare process for long rise times. These effects can make fast flares seem more powerful than they actually are.

There is another difference between these two types due to both the lengths of their rise times and their amplitudes. The lengths of rise times can reach to 1817 s for slow flares, but are not longer than 510 s for fast flares. In addition, when the flare amplitudes are examined for both types of flares, an adverse difference is seen contrary to rise times. While the amplitudes of slow flares reach to $0^{m}.480$ at most, the amplitudes of fast flares can exceed $1^{m}.430$.

Finally, computing the ratios of flare decay times to flare rise times for two types of flares, it is seen that the ratios do not exceed a specific value for all the slow flares. On the other hand, the ratios are always above this specific value for the fast flares. It means that the type of an observed flare can be determined by considering this value of the ratio. The limit value of this ratio is about 2.0 for the flares detected from \object{V1285 Aql}. However, \citet{Dal10} demonstrated that the limit values of the ratio of flare decay time to flare rise time is 3.5. Like the difference seen between the values of the equivalent durations given by \citet{Dal10} and the values of the equivalent durations given in this study for the fast and slow flares, the limit values of the ratios of flare decay times to flare rise times are also different. This must be again because of the same reasons.

Providing that the limit value of the ratios of flare decay times to flare rise times between flare types, the fast flare rate is 58.5$\%$ of the 83 flares observed in this study, while the slow flare rate is 41.5$\%$. It means that one of every two flares is the fast flare, other one is slow flare. This result diverges from what \citet{Gur88} stated. According to \citet{Gur88}, slow flares with low energies and low amplitudes make up 95$\%$ of all flares, and the remainders are fast flares.

It must be noted that comparing the correlation coefficients of the linear fits, it is seen that the correlation coefficient of the slow flares is quite higher than the fast flares. The same results was revealed by \citet{Dal10}. The difference is due to the equivalent durations of fast flares taking values in a wide range. Considering the non-thermal processes dominated in the fast flare events, \citet{Dal10} tried to explain it with the magnetic reconnection processes.

\subsection{The Saturation Level in the Detected White-Light Flare}

The distributions of flare-equivalent durations versus flare total duration were modelled by the OPEA function expressed by Equation (5) for 83 flares detected in U band observations of \object{V1285 Aql}. The regression calculations demonstrated that the best model function is the OPEA function for the distributions of flare-equivalent durations versus flare total duration. The derived model shows that flare-equivalent durations increase with the flare total duration until a specific total duration value, and then the flare-equivalent durations are constant no matter how long the flare total duration is.

According to the OPEA model of the flares detected from \object{V1285 Aql}, observed flare-equivalent durations start to reach maximum value in the value of 402.2 s of the flare total durations (note that the $Half-Life$ value is 201.1 s). The maximum value of the flare-equivalent durations is 2.421 s in logarithmic scales for the flare detected from \object{V1285 Aql}. It means that the flare processes occurring on the surface of \object{V1285 Aql} do not generate a flare more powerful than the specific value. This specific value of the flare-equivalent durations can be defined as a saturation level for the white-light flares occurring on the surface of \object{V1285 Aql}. The white-light flares occur in the regions, where the compact and two-ribbon flare events are seen \citep{Rod90, Ben10}. In the analyses, we used data obtained by the same method and the same optical system. In addition, we used the flare-equivalent durations instead of the flare energies. In fact, the derived $Plateau$ values depend only on the power of the white-light flares. Considering the $Plateau$ values, the flare-equivalent durations cannot be higher than a particular value no matter how long the flare total duration is. Instead of the flare duration, some other parameters, such as magnetic field flux and/or particle density in the volumes of the flare processes, must be more efficient in determining the power of the flares. Considering thermal and non-thermal flare events, both these parameters can be more efficient. However, Both \citet{Doy96a} and \citet{Doy96b} suggested that the saturation in the active stars does not have to be related to the filling factor of magnetic structures on the stellar surfaces or the dynamo mechanism under the surface. It can be related to some radiative losses in the chromosphere, where the temperature and density are increasing in the case of fast rotation. This phenomenon can occur in the chromosphere due to the flare process instead of fast rotation, and this causes the $Plateau$ phase to occur in the distributions of flare-equivalent duration versus flare total duration.

On the other hand, the $Plateau$ phase cannot be due to some radiative losses in the chromosphere with increasing temperature and density. This is because \citet{Gri83} demonstrated the effects of radiative losses in the chromosphere on the white-light photometry of the flares. According to \citet{Gri83}, the negative H opacity in the chromosphere causes the radiative losses, and these are seen as pre-flare dip in the light curves of the white-light flares. Unfortunately, considering the results of \citet{Dal11a}, it is seen that the $Plateau$ values vary from one star to the next. This indicates that some parameters or their efficacies, which make the $Plateau$ increase, are changing from star to star. It is seen that \object{V1285 Aql} is among its analogues. According to Standard Magnetic Reconnection Model developed by \citet{Pet64}, there are several important parameters giving shape to flare events, such as Alfv\'{e}n velocity ($\nu_{A}$), $B$, the emissivity of the plasma ($R$) and the most important one, the electron density of the plasma ($n_{e}$) \citep{VanB88, VanA88}. All these parameters are related with both heating and cooling processes in a flare event. \citet{VanA88, VanB88} have defined the radiative loss timescale ($\tau_{d}$) as $E_{th}/R$. Here $E_{th}$ is the total thermal energy, while $R$ is emissivity of the plasma. $E_{th}$ depends on the magnetic energy, which is defined as $B^{2}/8\pi$, and $R$ depends on the electron density ($n_{e}$) of the plasma. $\tau_{d}$ is firmly correlated with $B$ and $n_{e}$, while $\tau_{r}$ is proportional to a larger loop length ($\ell$) and smaller $B$ values. Consequently, it is seen that both the shape and power of a flare event depend on mainly two parameters, $n_{e}$ and $B$.

In addition, obtained maximum flare duration for \object{V1285 Aql} flares is 4641 s. This duration is in agreement with those found by \citet{Dal11a}. Observed maximum duration is 2940 s for \object{EV Lac}, and 3180 s for \object{EQ Peg}. The flares of both stars are dramatically lower than that of \object{V1285 Aql}. Maximum flare duration of \object{V1285 Aql} is almost 1.5 times of them. This case reveals some clues about the flaring loop geometry on these stars \citet{Ree02, Ima03, Fav05, Pan08}.

\subsection{Rotational Modulation and the Flare Distributions versus Rotational Phase}

\object{V1285 Aql} was observed in BVR bands apart from U band. Using the DFT method \citep{Sca82}, the data sets of BVR bands were separately  and together analysed for each observing season. It was found that the photometric period of the sinusoidal-like variation due to rotational modulation out-of-flare is $3^{d}.1269$ in B band, $3^{d}.1265$ in V band, $3^{d}.1268$ in R band. All the results obtained from the DFT method were tested by PDM and CLEANest methods. Examining the light variations, it is seen that the amplitudes are lower in some degree, but all the amplitudes are higher than the level of $3\sigma$. Besides, there are some variations in the colour curves, too. Considering the B-V index of \object{V1285 Aql}, the sinusoidal-like variation due to rotational modulation out-of-flare must be due to the heterogeneous temperature distribution on the surface of \object{V1285 Aql}. There must be some dark stellar spots on the surface. The variation of the V-R colour also supports this case. However, we do not see any variation above the level of $3\sigma$ in the B-V colour.

In this study, it was found that B-V colour index is $1^{m}.469$. It was given as $1^{m}.53$ by \citet{Pet91}, while it was given as $1^{m}.75$ by \citet{Mes01}. Different B-V indexes have been given in three studies for \object{V1285 Aql}. This could be due to both the methods and parameters used to find B-V index. If the methods and/or parameters are a little bit different, the obtained B-V can be slightly different. On the other hand, the main reason of these differences must be the magnetic activity seen on the surface of the star. The level of the flare activity observed on the star is very high. Although the same activity level is not seen in the light curves for the stellar rotational modulation, but all the surface of the star could be covered by dark stellar spots. If this is the case, it means that the level of the magnetic activity is rather higher than it is observed in the light curves out-of-flares. This caused to vary the observed B-V colour indexes from one study to next. Apart from B-V colour index of the star, found photometric period is also a bit different from the period found by \citet{Doy87}. The photometric period found in this study is about $3^{d}.127$, while the period found by \citet{Doy87} is $2^{d}.19$. This difference must be due to the variations of the locations of the spotted area(s). It is possible that the photometric period might be changing because of this.

Using the models shown with dashed lines in Figure 8, which were derived with analysing of each V band data set, the minimum, maximum, mean levels and the amplitudes of the light curves were computed. Examining these parameters listed in Table 5, the levels of the brightness slowly varies season to season. The variations are slow. This must be because of the small developing of the structures on the surface of the star.

There are many studies about whether the flares of UV Ceti type stars, which exhibit BY Dra Syndrome, are occurring at the same longitudes of stellar spots, or not \citep{Dal11b}. Having the same longitudes of flare and spots is an expected case for these stars, because solar flares are mostly occurring in the active regions, where spots are located on the Sun \citep{Ben10}. In the respect of Stellar-Solar Connection, a result of the $Ca$ $II$ $H\&K$ Project of Mount Wilson Observatory \citep{Wil78, Bal95}, if the areas of flares and spots are related on the Sun, the same case might be expected for the stars. In fact, \citet{Mon96} found some evidence to demonstrate this relation. Besides, \citet{Let97} found a variation of both the rotational modulation and the phase distribution of flare occurrence rates in the same way for the observations in the year 1970. On the other hand, no clear relation between stellar flares and spots was found by \citet{Bop74, Pet83} and \citet{Dal11b}. However, \citet{Pet83} did not draw firm conclusions because of being a non-uniqueness problem.

In this study, using the method described in Section 2.5, we derived the distribution of flares versus rotational phase for all the flares detected in the observing season 2006. The derived distribution is shown in Figure 9. According to the distribution, the phase of MFOR is seen between $0^{P}.40$ and $0^{P}.50$. Almost 6 flares were detected per hour in this phase interval, while 2 flares were detected per hour at most in all other phase intervals. This is a more important result, because this phase interval is where the minimum phase of the sinusoidal-like variation due to rotational modulation out-of-flare is seen in the V band light curves of the observing season 2006. The phase of the sinusoidal-like variation is $0^{P}.47$.

Using the inverse Compton event, \citet{Gur86} developed the Fast Electron Hypothesis. According to this hypothesis, UV Ceti type stars should generate only fast flares on their surface. However, the shapes of the flare light variations can be seen like a slow flare in respect to direction of observer \citep{Gur86}. According to the Fast Electron Hypothesis, it should be expected that both the fast and slow flares get groups, which are separated $0^{P}.50$ from each other. In this study, using the method described by \citet{Dal11b}, 48 fast and 35 slow flares were defined and their phases were computed. As seen from the analyses, the slow flares occur more frequently around $0^{P}.30$, while the fast flares are occurring more frequently around $0^{P}.45$. Unfortunately, the difference between the phases of two flare types is $0^{P}.15$ instead of $0^{P}.50$. Apart from this unexpected value, both the fast and the slow flares occur almost all phase intervals.

In conclusion, some parameters can be computed from flares observed in photoelectric photometry and if the behaviours between these parameters can be analysed by suitable methods, flare types can be determined. In this study, we analysed the distributions of equivalent durations versus flare rise times by using a t-Test. Finally, it is seen that using the ratios of flare decay times to flare rise times, flares can be classified. It is seen that there are considerable differences between these two types of flares. Moreover, analyses demonstrated the detected flares have some critical energy level. There is no flare, whose energy is much more than this level. In addition, analyses demonstrated that \object{V1285 Aql} exhibits stellar spot activity apart from flares. Comparing the phase distribution of the flares with the phase of the sinusoidal-like variation demonstrated that the flares have a tendency to occur in the same longitudes with stellar spots. On the other hand, according to the statistical analyses, the slow flares and fast flares are not separated with some definite rules from each other. In this respect, extending the B-V range of observed stars, required to obtain more data, which should be obtained from many different stars and flare patrols spanning many years, in order to obtain more reliable results.



\acknowledgments

The authors acknowledge generous allotments of observing time at the Ege University Observatory. We thank the staffs working at EUO. We thank the referee for useful comments that have contributed to the improvement of the paper. We also thank Professor Dr. M. Can Akan who gave us valuable suggestions that improved the language of the manuscript.



{\it Facilities:} \facility{HSTCP}, \facility{A48}, \facility{Ege University Observatory (EUO)}.


\clearpage



\begin{figure}
\epsscale{1.00}
\plotone{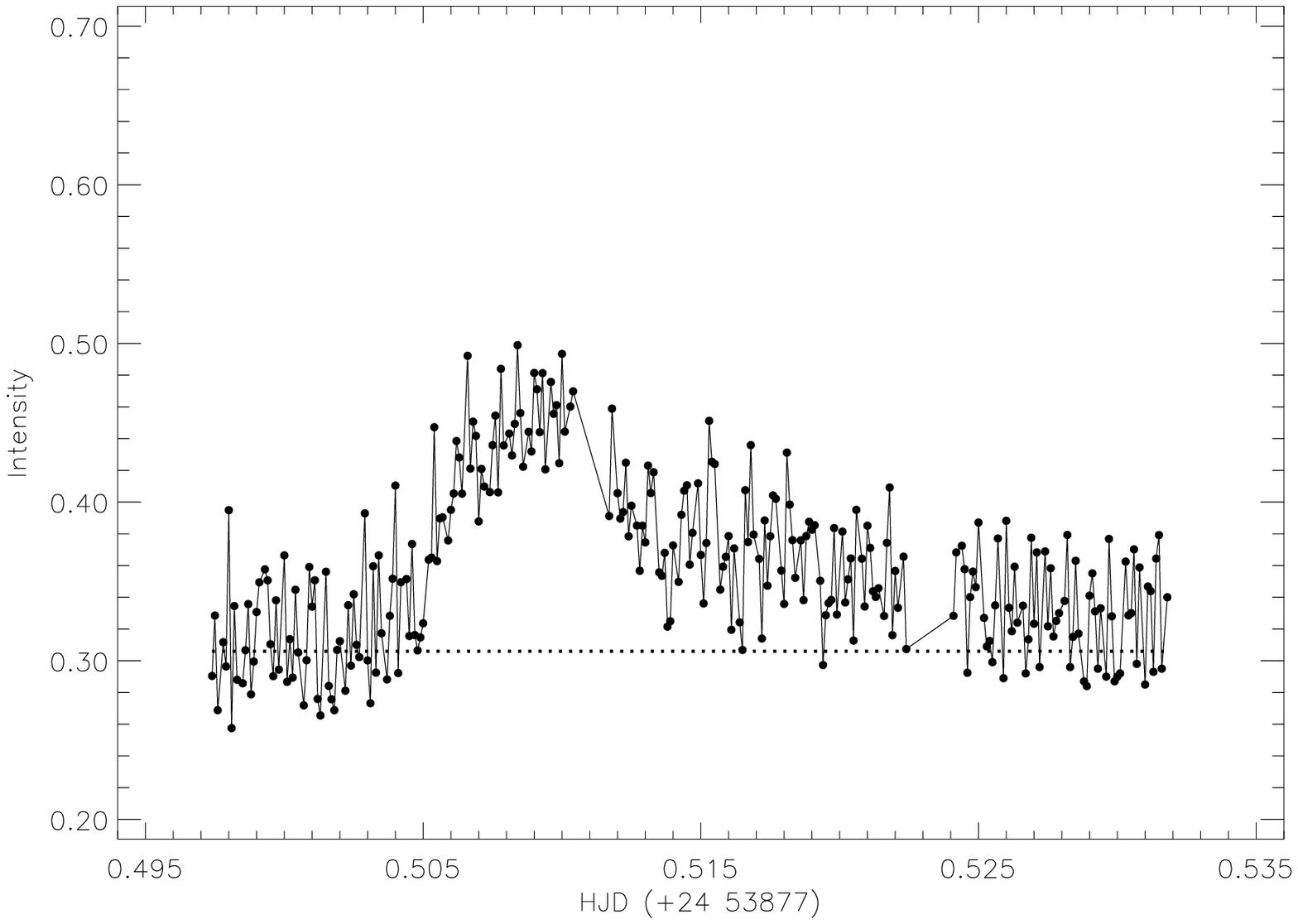}
\caption{A flare light curve sample for fast flares obtained from U-band observations of V1285 Aql on 2006, May 21.\label{fig1}}
\end{figure}

\begin{figure}
\epsscale{1.00}
\plotone{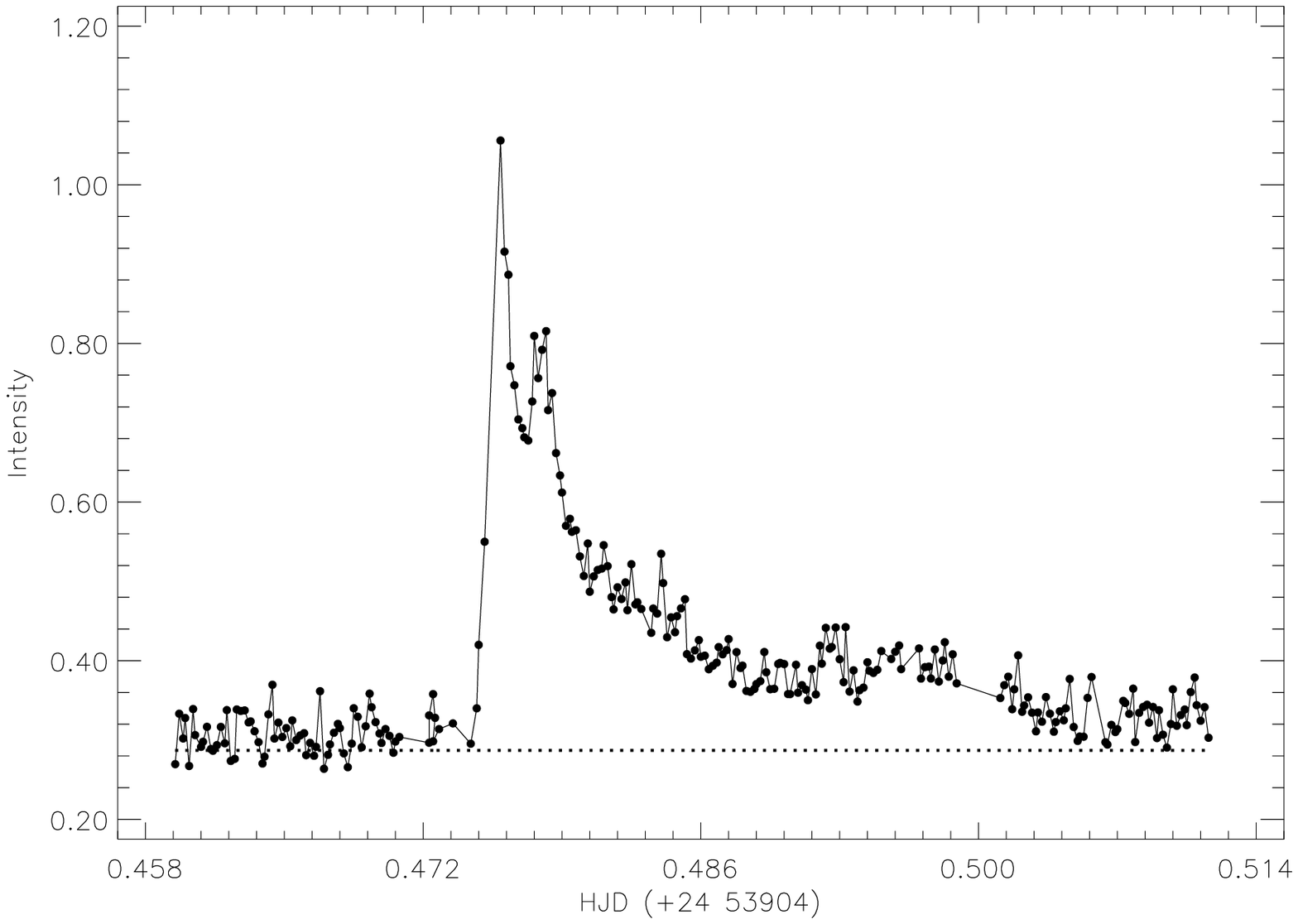}
\caption{A flare light curve sample for fast flares obtained from U-band observations of V1285 Aql on 2006, June 17.\label{fig2}}
\end{figure}

\begin{figure}
\epsscale{1.00}
\plotone{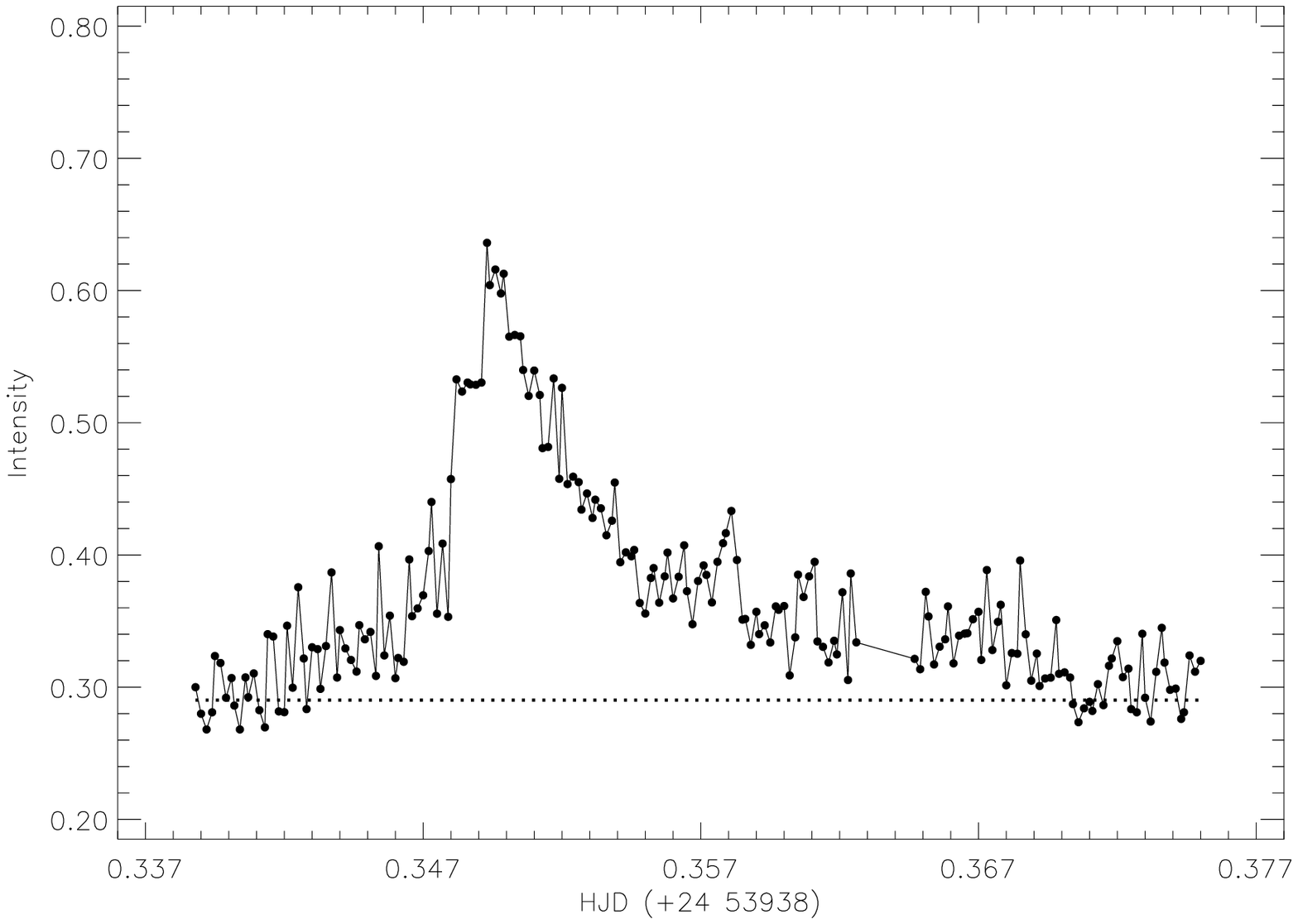}
\caption{A flare light curve sample for fast flares obtained from U-band observations of V1285 Aql on 2006, July 21.\label{fig3}}
\end{figure}

\begin{figure}
\epsscale{1.00}
\plotone{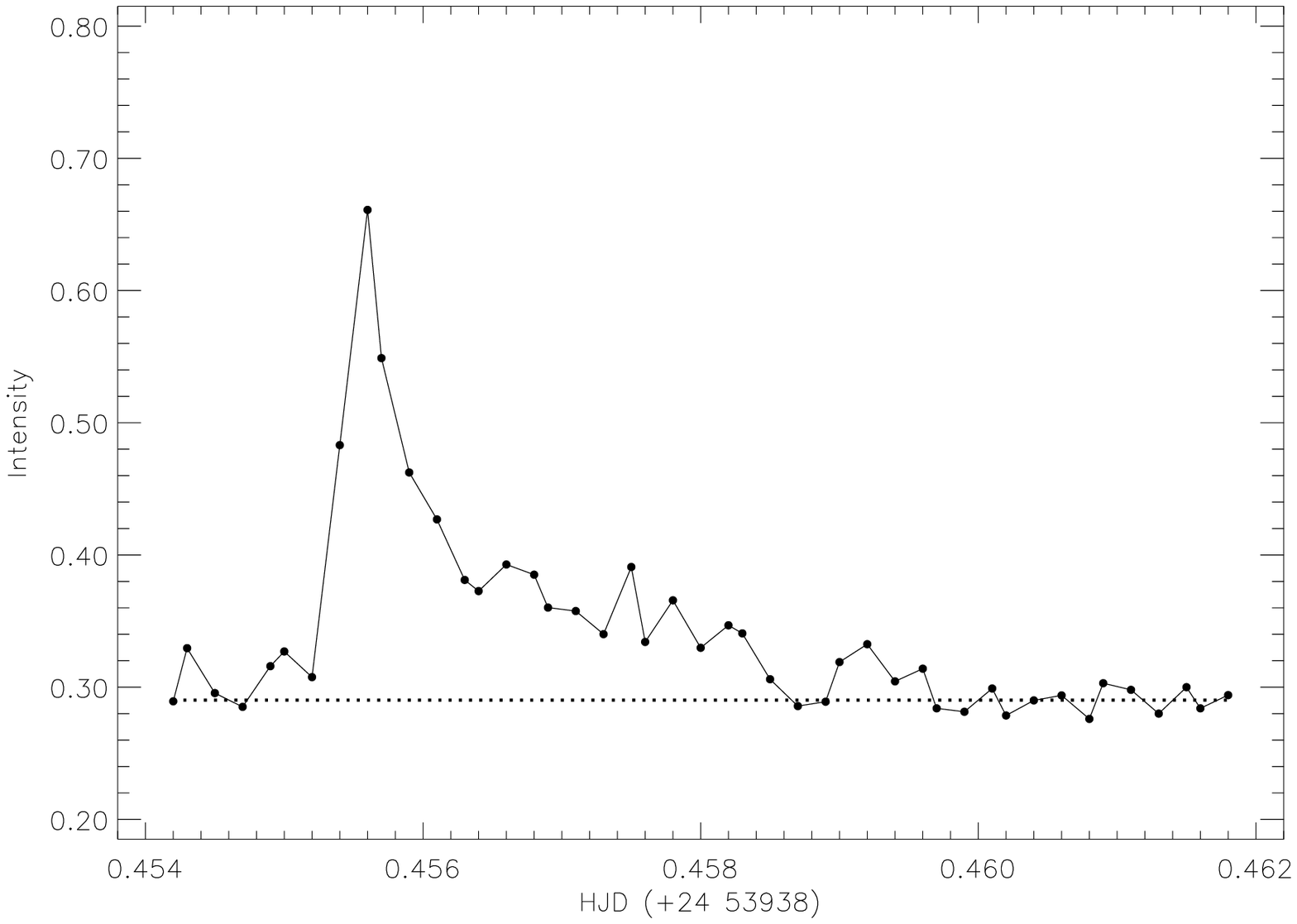}
\caption{A flare light curve sample for fast flares obtained from U-band observations of V1285 Aql on 2006, July 21.\label{fig4}}
\end{figure}

\begin{figure}
\epsscale{1.00}
\plotone{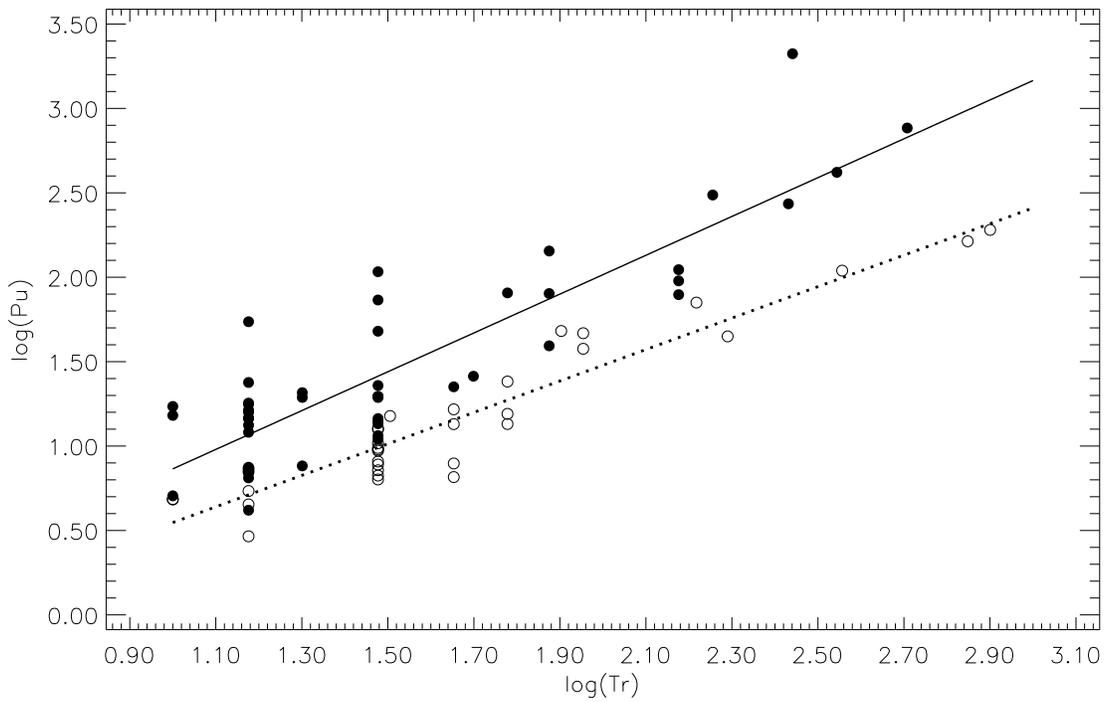}
\caption{Distributions for the mean averages of the equivalent durations ($logP_{u}$) vs. flare rise times ($logT_{r}$) in the logarithmic scale. In the figure, open circles represent slow flares, while filled circles show the fast flares. Lines represent fits given in Equations (3) and (4).\label{fig5}}
\end{figure}

\begin{figure}
\epsscale{1.00}
\plotone{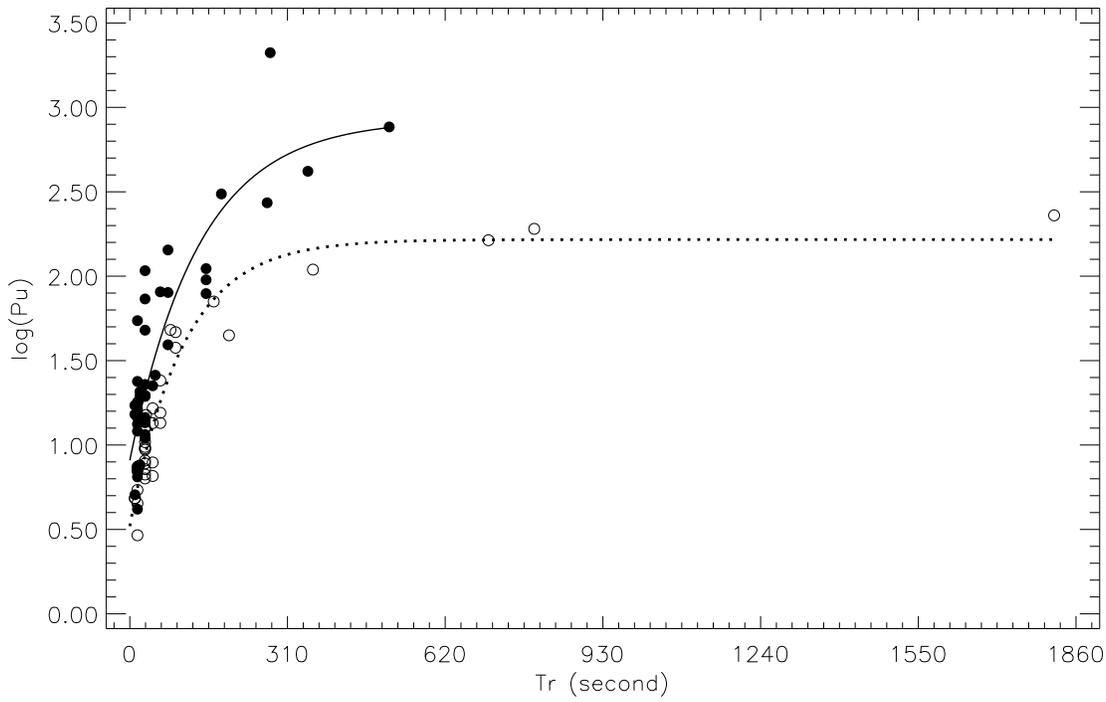}
\caption{ Distributions of the equivalent durations ($logP_{u}$) in the logarithmic scale vs. flare rise times ($T_{r}$) for all 83 flares detected in observations of program stars. In the figure open circles represent slow flares, while filled circles show the fast flares.\label{fig6}}
\end{figure}

\begin{figure}
\epsscale{1.00}
\plotone{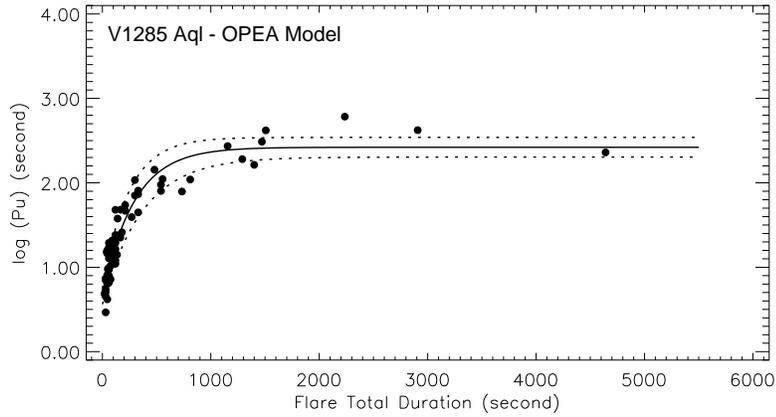}
\caption{Distributions of flare-equivalent duration on a logarithmic scale vs. flare total duration. Filled circles represent equivalent durations computed from flares detected from V1285 Aql. The line represents the model identified with Equation (5) computed using the least-squares method. The dotted lines represent 95$\%$ confidence intervals for the model.\label{fig7}}
\end{figure}

\begin{figure}
\epsscale{1.00}
\plotone{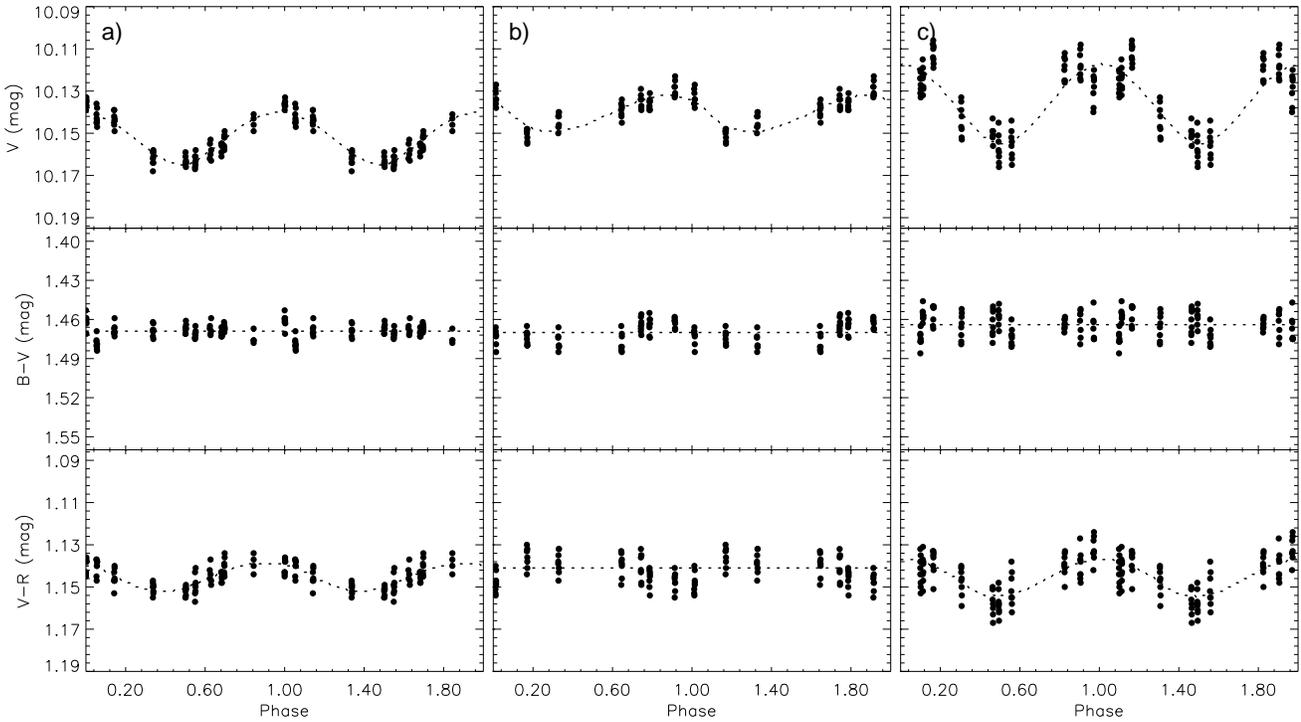}
\caption{V band light and B-V and V-R colour curves obtained in this study are seen for three data sets composed from the observations of V1285 Aql. a) Observing season 2006, b) Observing season 2007, c) Observing season 2008. In the panels of the light curves, dashed lines represent the fits derived from the Discrete Fourier Transform.\label{fig8}}
\end{figure}

\begin{figure}
\epsscale{1.00}
\plotone{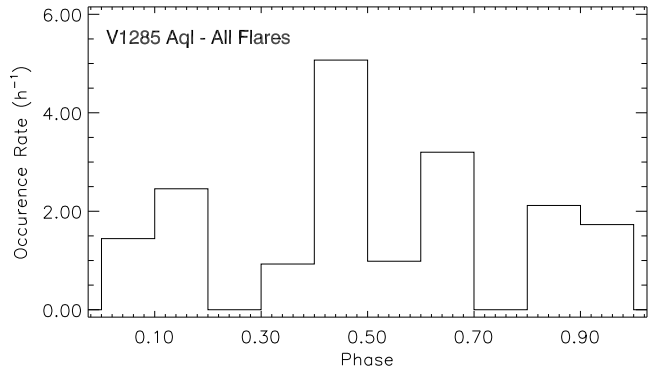}
\caption{The mean flare occurrence rates versus rotational phase are demonstrated for all V1285 Aql flares observed in the season 2006. In the figure, the line shows the histogram of mean flare occurrence rates computed in intervals of 0.10 phase length. All 83 flares (combining the fast and slow flares together) were counted in the calculation.\label{fig9}}
\end{figure}

\begin{figure}
\epsscale{1.00}
\plotone{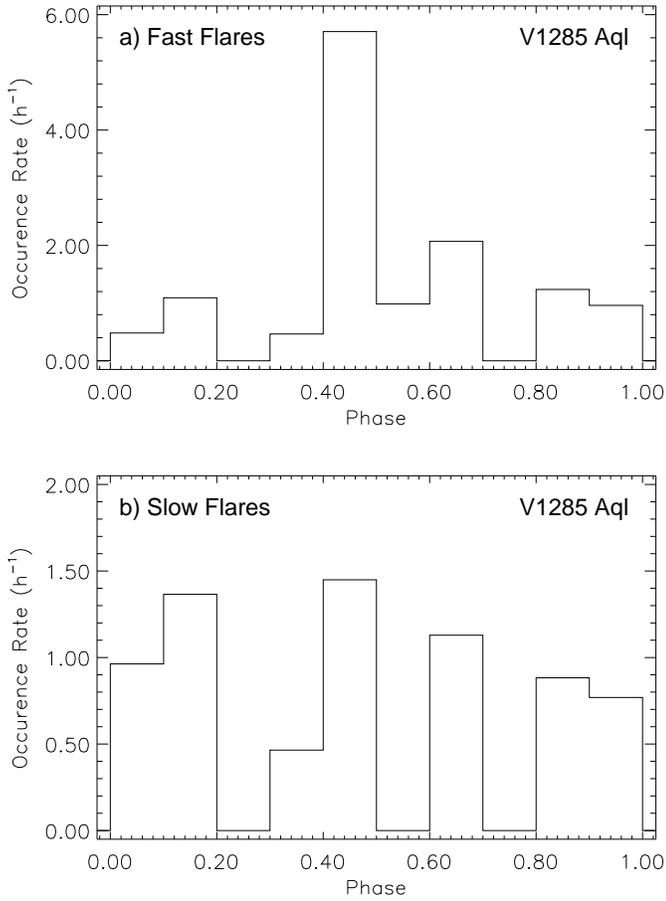}
\caption{The mean flare occurrence rates versus rotational phase are demonstrated for the V1285 Aql flares observed in the season 2006. In the figures, the lines show the histograms of mean flare occurrence rates computed in intervals of 0.10 phase length. The histogram of the fast flares is shown in upper panel, while it is shown for the slow flares in bottom panel.\label{fig10}}
\end{figure}







\clearpage

\begin{table}
\begin{center}
\caption{Basic parameters for the star studied and its comparison (C1) and check (C2) stars.\label{tbl-1}}
\begin{tabular}{lcc}
\hline\hline
\textbf{Stars}	&	\textbf{V (mag)}	&	\textbf{B-V (mag)}	\\
\hline					
V1285 Aql	&	10.142	&	1.469	\\
C1 = BD +08 3899	&	9.651	&	1.137	\\
C2 = BD +08 3900	&	9.994	&	1.410	\\
\hline
\end{tabular}
\end{center}
\end{table}

\begin{deluxetable}{cccccccccc}
\tabletypesize{\scriptsize}
\tablecaption{The parameters derived from analyses of the detected flares.\label{tbl-2}}
\tablewidth{0pt}
\tablehead{
\colhead{Observing}	&	\colhead{HJD of Flare} & \colhead{Rise} &	\colhead{Decay} &	\colhead{Total} &	\colhead{Equivalent}	&	\colhead{Flare}	&	\colhead{Flare}	&	\colhead{Flare}	&	\colhead{Flare}	\\
\colhead{} &	\colhead{Maximum}	& \colhead{Time}	&	\colhead{Time} &	\colhead{Duration} & \colhead{Duration}	& \colhead{Amplitude}	& \colhead{U-B}	& \colhead{Energy}	&	\colhead{}	\\
 \colhead{Date}	&	\colhead{(+24 00000)}	&	\colhead{(s)}	&	\colhead{(s)}	&	\colhead{(s)}	&	\colhead{(s)}	&	\colhead{(mag)}	&	\colhead{(mag)}	&	\colhead{(erg)}	&	\colhead{Type}
}
\startdata																			
19.05.2006	&	53875.53888	&	32	&	32	&	64	&	15.06684	&	0.357	&	0.095	&	7.10348E+30	&	Slow	\\
21.05.2006	&	53877.47941	&	30	&	20	&	50	&	9.54445	&	0.477	&	0.073	&	4.49987E+30	&	Slow	\\
21.05.2006	&	53877.50898	&	350	&	1157	&	1507	&	418.16843	&	0.462	&	-0.043	&	1.97152E+32	&	Fast	\\
26.05.2006	&	53882.46588	&	10	&	50	&	60	&	17.16708	&	0.353	&	0.122	&	8.09367E+30	&	Fast	\\
26.05.2006	&	53882.46901	&	30	&	80	&	110	&	19.68734	&	0.259	&	0.203	&	9.28188E+30	&	Fast	\\
26.05.2006	&	53882.47190	&	10	&	10	&	20	&	4.82332	&	0.417	&	0.057	&	2.27402E+30	&	Slow	\\
26.05.2006	&	53882.48450	&	10	&	20	&	30	&	5.07340	&	0.289	&	0.170	&	2.39193E+30	&	Fast	\\
26.05.2006	&	53882.48902	&	20	&	40	&	60	&	7.63124	&	0.359	&	0.147	&	3.59786E+30	&	Fast	\\
30.05.2006	&	53886.36985	&	20	&	70	&	90	&	20.72296	&	0.341	&	0.065	&	9.77014E+30	&	Fast	\\
30.05.2006	&	53886.37100	&	30	&	70	&	100	&	19.67298	&	0.230	&	0.268	&	9.27511E+30	&	Fast	\\
30.05.2006	&	53886.37228	&	10	&	10	&	20	&	4.85286	&	0.438	&	0.080	&	2.28795E+30	&	Slow	\\
30.05.2006	&	53886.37575	&	10	&	30	&	40	&	15.19447	&	0.370	&	0.033	&	7.16365E+30	&	Fast	\\
30.05.2006	&	53886.37887	&	20	&	40	&	60	&	19.41469	&	0.317	&	0.116	&	9.15334E+30	&	Fast	\\
30.05.2006	&	53886.38026	&	80	&	90	&	170	&	48.04427	&	0.461	&	-0.041	&	2.26512E+31	&	Slow	\\
30.05.2006	&	53886.40733	&	90	&	50	&	140	&	37.68882	&	0.453	&	0.059	&	1.77689E+31	&	Slow	\\
30.05.2006	&	53886.41347	&	50	&	130	&	180	&	25.92122	&	0.274	&	0.210	&	1.22209E+31	&	Fast	\\
17.06.2006	&	53904.37813	&	180	&	1291	&	1471	&	306.95989	&	0.282	&	-0.072	&	1.44721E+32	&	Fast	\\
17.06.2006	&	53904.39776	&	30	&	45	&	75	&	12.57656	&	0.328	&	0.197	&	5.92940E+30	&	Slow	\\
17.06.2006	&	53904.39984	&	60	&	30	&	90	&	15.51715	&	0.350	&	0.167	&	7.31579E+30	&	Slow	\\
17.06.2006	&	53904.40366	&	15	&	30	&	45	&	7.49366	&	0.281	&	0.151	&	3.53300E+30	&	Fast	\\
17.06.2006	&	53904.40922	&	195	&	135	&	330	&	44.64127	&	0.241	&	0.252	&	2.10468E+31	&	Slow	\\
17.06.2006	&	53904.41338	&	30	&	45	&	75	&	10.39874	&	0.251	&	0.215	&	4.90264E+30	&	Slow	\\
17.06.2006	&	53904.46263	&	15	&	105	&	120	&	12.07430	&	0.221	&	0.336	&	5.69260E+30	&	Fast	\\
17.06.2006	&	53904.46437	&	30	&	75	&	105	&	11.53398	&	0.305	&	0.230	&	5.43786E+30	&	Fast	\\
17.06.2006	&	53904.46680	&	15	&	15	&	30	&	2.92073	&	0.234	&	0.228	&	1.37702E+30	&	Slow	\\
17.06.2006	&	53904.46923	&	30	&	105	&	135	&	14.06378	&	0.227	&	0.283	&	6.63057E+30	&	Fast	\\
17.06.2006	&	53904.47131	&	30	&	90	&	120	&	10.98540	&	0.204	&	0.277	&	5.17923E+30	&	Fast	\\
17.06.2006	&	53904.47589	&	276	&	2632	&	2908	&	2107.73499	&	1.437	&	-0.745	&	9.93722E+32	&	Fast	\\
18.06.2006	&	53905.33006	&	165	&	135	&	300	&	70.78081	&	0.411	&	0.093	&	3.33706E+31	&	Slow	\\
18.06.2006	&	53905.33474	&	270	&	885	&	1155	&	272.17565	&	0.397	&	0.135	&	1.28321E+32	&	Fast	\\
18.06.2006	&	53905.35978	&	15	&	30	&	45	&	7.19677	&	0.349	&	0.160	&	3.39302E+30	&	Fast	\\
18.06.2006	&	53905.36047	&	30	&	90	&	120	&	22.84462	&	0.254	&	0.244	&	1.07704E+31	&	Fast	\\
18.06.2006	&	53905.36204	&	15	&	90	&	105	&	13.29829	&	0.291	&	0.216	&	6.26967E+30	&	Fast	\\
18.06.2006	&	53905.36796	&	30	&	30	&	60	&	9.75576	&	0.318	&	0.168	&	4.59949E+30	&	Slow	\\
18.06.2006	&	53905.37533	&	15	&	30	&	45	&	15.85617	&	0.354	&	0.153	&	7.47562E+30	&	Fast	\\
18.06.2006	&	53905.38783	&	795	&	495	&	1290	&	190.68324	&	0.190	&	0.229	&	8.99004E+31	&	Slow	\\
18.06.2006	&	53905.39862	&	15	&	60	&	75	&	17.70708	&	0.387	&	0.076	&	8.34826E+30	&	Fast	\\
18.06.2006	&	53905.40060	&	30	&	30	&	60	&	12.71769	&	0.258	&	0.181	&	5.99594E+30	&	Slow	\\
21.06.2006	&	53908.33621	&	30	&	90	&	120	&	14.55190	&	0.299	&	0.153	&	6.86070E+30	&	Fast	\\
21.06.2006	&	53908.33743	&	15	&	105	&	120	&	16.35125	&	0.241	&	0.158	&	7.70903E+30	&	Fast	\\
21.06.2006	&	53908.33899	&	15	&	45	&	60	&	6.98155	&	0.331	&	0.196	&	3.29155E+30	&	Fast	\\
21.06.2006	&	53908.33986	&	30	&	30	&	60	&	9.38434	&	0.299	&	0.154	&	4.42438E+30	&	Slow	\\
21.06.2006	&	53908.34055	&	15	&	30	&	45	&	7.28724	&	0.288	&	0.189	&	3.43567E+30	&	Fast	\\
21.06.2006	&	53908.34107	&	15	&	15	&	30	&	5.41634	&	0.314	&	0.134	&	2.55361E+30	&	Slow	\\
21.06.2006	&	53908.34142	&	15	&	15	&	30	&	4.51176	&	0.395	&	0.092	&	2.12713E+30	&	Slow	\\
21.06.2006	&	53908.34194	&	15	&	30	&	45	&	4.16976	&	0.268	&	0.201	&	1.96589E+30	&	Fast	\\
21.06.2006	&	53908.34368	&	30	&	15	&	45	&	6.69016	&	0.390	&	0.088	&	3.15417E+30	&	Slow	\\
21.06.2006	&	53908.34437	&	15	&	90	&	105	&	14.64679	&	0.289	&	0.166	&	6.90544E+30	&	Fast	\\
21.06.2006	&	53908.34559	&	15	&	15	&	30	&	7.04872	&	0.349	&	0.095	&	3.32322E+30	&	Slow	\\
21.06.2006	&	53908.34732	&	60	&	30	&	90	&	13.51126	&	0.296	&	0.202	&	6.37008E+30	&	Slow	\\
21.06.2006	&	53908.34819	&	45	&	15	&	60	&	6.55336	&	0.372	&	0.173	&	3.08968E+30	&	Slow	\\
21.06.2006	&	53908.37044	&	75	&	405	&	480	&	142.99419	&	0.728	&	-0.210	&	6.74167E+31	&	Fast	\\
21.06.2006	&	53908.44178	&	1817	&	2824	&	4641	&	229.22026	&	0.171	&	-0.022	&	2.30310E+32	&	Slow	\\
21.06.2006	&	53908.48645	&	150	&	405	&	555	&	110.87499	&	0.434	&	0.066	&	5.22736E+31	&	Fast	\\
21.06.2006	&	53908.49322	&	75	&	195	&	270	&	39.24736	&	0.368	&	0.154	&	1.85037E+31	&	Fast	\\
27.06.2006	&	53914.35370	&	360	&	450	&	810	&	109.39455	&	0.293	&	0.319	&	5.15756E+31	&	Slow	\\
27.06.2006	&	53914.37720	&	90	&	120	&	210	&	46.60488	&	0.322	&	0.242	&	2.19725E+31	&	Slow	\\
27.06.2006	&	53914.41381	&	75	&	465	&	540	&	80.08406	&	0.363	&	0.239	&	3.77568E+31	&	Fast	\\
27.06.2006	&	53914.45440	&	150	&	390	&	540	&	95.19813	&	0.332	&	0.212	&	4.48825E+31	&	Fast	\\
28.06.2006	&	53915.45427	&	15	&	135	&	150	&	23.80623	&	0.464	&	0.140	&	1.12238E+31	&	Fast	\\
28.06.2006	&	53915.46989	&	705	&	695	&	1400	&	163.30309	&	0.390	&	0.157	&	7.69916E+31	&	Slow	\\
21.07.2006	&	53938.34925	&	510	&	1726	&	2236	&	765.66892	&	0.821	&	-0.234	&	3.60986E+32	&	Fast	\\
21.07.2006	&	53938.44548	&	15	&	90	&	105	&	17.96942	&	0.402	&	0.122	&	8.47195E+30	&	Fast	\\
21.07.2006	&	53938.45556	&	30	&	270	&	300	&	107.79387	&	0.831	&	-0.323	&	5.08210E+31	&	Fast	\\
23.07.2006	&	53940.34400	&	15	&	30	&	45	&	6.47911	&	0.304	&	0.249	&	3.05467E+30	&	Fast	\\
23.07.2006	&	53940.35286	&	30	&	45	&	75	&	7.19161	&	0.381	&	0.230	&	3.39059E+30	&	Slow	\\
23.07.2006	&	53940.35563	&	150	&	584	&	734	&	78.86360	&	0.355	&	0.141	&	3.71814E+31	&	Fast	\\
23.07.2006	&	53940.36430	&	30	&	90	&	120	&	13.56160	&	0.296	&	0.140	&	6.39382E+30	&	Fast	\\
23.07.2006	&	53940.36708	&	30	&	15	&	45	&	6.34242	&	0.251	&	0.143	&	2.99023E+30	&	Slow	\\
23.07.2006	&	53940.37350	&	30	&	30	&	60	&	7.77530	&	0.220	&	0.182	&	3.66578E+30	&	Slow	\\
23.07.2006	&	53940.37559	&	45	&	30	&	75	&	13.49058	&	0.263	&	0.232	&	6.36033E+30	&	Slow	\\
27.07.2006	&	53944.34396	&	15	&	195	&	210	&	54.53049	&	0.368	&	0.213	&	2.57092E+31	&	Fast	\\
27.07.2006	&	53944.34743	&	60	&	60	&	120	&	24.11020	&	0.388	&	0.244	&	1.13671E+31	&	Slow	\\
27.07.2006	&	53944.34951	&	30	&	300	&	330	&	73.31641	&	0.366	&	0.324	&	3.45661E+31	&	Fast	\\
27.07.2006	&	53944.35368	&	15	&	30	&	45	&	14.56958	&	0.410	&	0.185	&	6.86904E+30	&	Fast	\\
27.07.2006	&	53944.36449	&	60	&	270	&	330	&	80.76671	&	0.517	&	0.053	&	3.80786E+31	&	Fast	\\
27.07.2006	&	53944.37213	&	30	&	90	&	120	&	47.89915	&	0.574	&	0.098	&	2.25828E+31	&	Fast	\\
29.07.2006	&	53946.36976	&	45	&	60	&	105	&	16.50538	&	0.288	&	0.231	&	7.78170E+30	&	Slow	\\
29.07.2006	&	53946.42235	&	45	&	15	&	60	&	7.88789	&	0.291	&	0.179	&	3.71886E+30	&	Slow	\\
29.07.2006	&	53946.42373	&	45	&	120	&	165	&	22.43408	&	0.277	&	0.233	&	1.05769E+31	&	Fast	\\
29.07.2006	&	53946.42651	&	30	&	15	&	45	&	8.10063	&	0.344	&	0.191	&	3.81916E+30	&	Slow	\\
02.08.2006	&	53950.34266	&	30	&	90	&	120	&	19.39572	&	0.352	&	0.133	&	9.14439E+30	&	Fast	\\
02.08.2006	&	53950.35061	&	15	&	15	&	30	&	7.37094	&	0.328	&	0.147	&	3.47514E+30	&	Slow	\\
\enddata
\end{deluxetable}

\clearpage

\begin{table*}
\begin{center}
\caption{For both fast and slow flares whose rise times are the same. The results obtained from both the regression calculations and the t-Test analyses performed to the mean averages of the equivalent durations ($logP_{u}$) versus flare rise times ($logT_{r}$) in the logarithmic scale are listed.\label{tbl-3}}
\begin{tabular}{lrr}
\tableline\tableline					
\textbf{Flare Groups :} 	&	 \textbf{Slow Flare} 	&	 \textbf{Fast Flare} 	\\
\tableline									
\textbf{\textit{Best Representation Values}}	&		&		\\				
Slope : 	&	0.932$\pm$0.056	&	1.150$\pm$0.095	\\
$y-intercept$ when $x = 0.0$ : 	&	-0.385$\pm$0.096	&	-0.285$\pm$0.151	\\
$x-intercept$ when $y = 0.0$ : 	&	0.414	&	0.248	\\
\tableline\tableline									
\textbf{\textit{Mean Average of All Y Values}} 	&	 	&	 	\\				
Mean Average : 	&	1.479	&	2.015	\\
Mean Average Error : 	&	0.054	&	0.067	\\
\tableline\tableline									
\textbf{\textit{Goodness of Fit}} 	&	 	&	 	\\				
$r^{2}$ : 	&	0.896	&	0.761	\\
\tableline\tableline									
\textbf{\textit{Is slope significantly non-zero?}} 	&	 	&	 	\\				
$p-value$ : 	&	 $<$ 0.0001 	&	$<$ 0.0001	\\
Deviation from zero? : 	&	 $Significant$ 	&	$Significant$	\\
\tableline
\end{tabular}
\end{center}
\end{table*}

\begin{table}
\begin{center}
\caption{Using the Least-Squares Method, the parameters were obtained from the OPEA function.\label{tbl-4}}
\begin{tabular}{lrr}
\tableline\tableline
\textbf{Parameter}	&	\textbf{Value}	&	\textbf{Error}	\\
\tableline					
$y_{0}$	&	0.656	&	0.048	\\
$Plateau$	&	2.421	&	0.058	\\
$k$	&	0.003447	&	0.000365	\\
$Tau$	&	290.1	&	-	\\
$Half-Life$	&	201.1	&	-	\\
$Span$	&	1.765	&	0.064	\\
\tableline
\end{tabular}
\end{center}
\end{table}

\begin{table*}
\begin{center}
\caption{The maximum, minimum, mean brightness levels and amplitudes of V band light curves obtained with using the light elements given in Equation (6), which was found by the DFT method \citet{Sca82}.\label{tbl-5}}
\begin{tabular}{ccccc}
\tableline\tableline
Observing	&	$V_{min}$	&	$V_{max}$	&	$V_{mean}$	&	Amplitude	\\
Season	&	(mag)	&	(mag)	&	(mag)	&	(mag)	\\
\tableline									
2006	&	10.164	&	10.137	&	10.152	&	0.027	\\
2007	&	10.152	&	10.128	&	10.139	&	0.024	\\
2008	&	10.158	&	10.115	&	10.135	&	0.043	\\
\tableline
\end{tabular}
\end{center}
\end{table*}




\end{document}